\documentclass{ws-procs9x6}            % default, citations in superscript
\usepackage{multirow}
\usepackage[colorlinks,citecolor=blue,urlcolor=blue,linkcolor=blue]{hyperref}
 \usepackage{indentfirst}
\begin{document}
\title{Hidden charm pentaquark states and $\Sigma_c^{(*)}\bar{D}^{(*)}$ interaction in chiral perturbation theory }

\author{Lu Meng$^*$, Bo Wang, Guang-Juan Wang and Shi-Lin Zhu}

\address{School of Physics and State Key Laboratory of Nuclear
	Physics and Technology, Peking University, Beijing 100871,
	China\\
$^*$E-mail: lmeng@pku.edu.cn}

\begin{abstract}
	We adopt the chiral perturbation theory to calculate the $\Sigma_{c}^{(*)}\bar{D}^{(*)}$ interaction to the next-to-leading order (NLO) and include the couple-channel effect in the loop diagrams. We reproduce the three $P_c$ states in the molecular picture after including the $\Lambda_{c}\bar{D}^{(*)}$ intermediate states. We also discuss some novel observations arising from the loop diagrams.
\end{abstract}


\bodymatter

\section{Introduction}\label{aba:sec1}

In 2015, the LHCb Collaboration discovered two pentaquark candidates $P_c(4380)$ and $P_c(4450)$ in the $J/\psi p$ invariant mass spectrum of $\Lambda_b\rightarrow J/\psi Kp$~\cite{Aaij:2015tga} (or see Ref.~\citenum{Chen:2016qju} for a review). Recently, the LHCb Collaboration updated these results~\cite{Aaij:2019vzc}. The previously reported $P_c(4450)$ was resolved into two states $P_c(4440)$ and $P_c(4457)$, and a new state $P_c(4312)$ was observed with 7.3$\sigma$ significance. These three states are narrow and their masses lie below the thresholds of $\Sigma_c\bar{D}^{(*)}$. Thus, they are good candidates of the $\Sigma_c\bar{D}^{(*)}$ molecules. The theoretical calculation in different approaches all supports their molecular nature~\cite{Chen:2019asm,Liu:2019tjn,Guo:2019fdo,Du:2019pij}.

Chiral perturbation theory (ChPT) is the effective field theory of low energy QCD, which was used to build the modern theory of nuclear force~\cite{Epelbaum:2008ga,Machleidt:2011zz}. In this work, we adopt the ChPT to study the $\Sigma_{c}^{(*)}\bar{D}^{(*)}$ interaction~\cite{Meng:2019ilv,Wang:2019ato}, which is less model-dependent than other previous works. We take the loop diagrams into consideration, which give rise to some novel effects.

\section{Formalism}
\begin{figure}
	\begin{center}
		\includegraphics[width=1.0\textwidth]{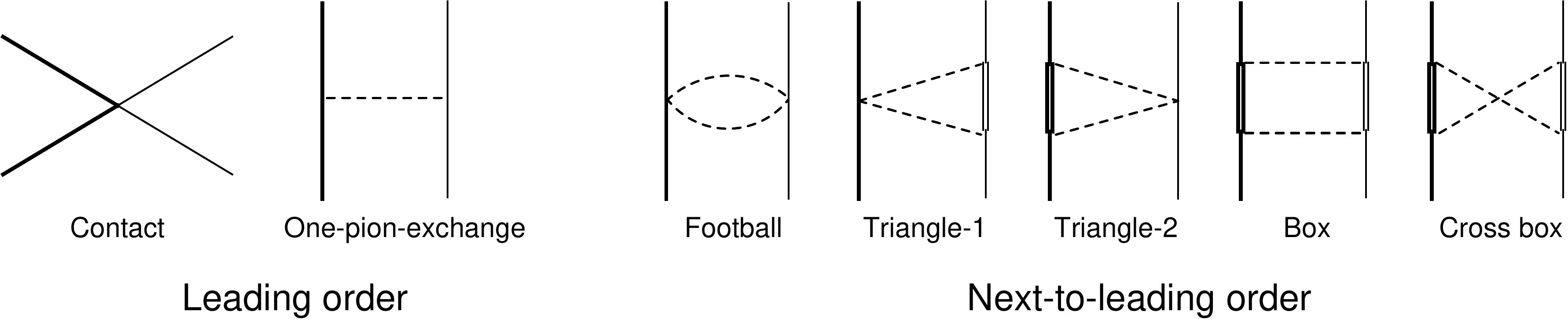}
	\end{center}
	\caption{Topological diagrams contributing to the LO and NLO $\Sigma_{c}^{(*)}\bar{D}^{*}$ interaction.}
	\label{fig:topo}
\end{figure}

The ChPT is performed to calculate the $\Sigma_{c}^{(*)}\bar{D}^{*}$ effective potentials to the next-to-leading order (NLO). The topological diagrams involved in the calculation are presented in Fig.~\ref{fig:topo}. The leading order (LO) diagrams include the tree level contact interaction and one-pion-exchange diagrams. The NLO diagrams are the two-pion-exchange ones. In our calculation, the intermediate matter fields (double lines in Fig.~\ref{fig:topo}) could be $\Sigma_{c}$, $\Sigma_{c}^*$, $\bar{D}$ and $\bar{D}^*$. We keep the mass splittings between the heavy quark spin (HQS) partner states, thus our results could give the heavy quark spin symmetry (HQSS) violation effect. Since the $\Sigma_c^{(*)}\Lambda_{c}\pi$ couplings are competitive compared with the $\Sigma_c^{(*)}\Sigma_c^{(*)}\pi$ ones, we also include the $\Lambda_{c}$ as the intermediate states in the loops.

\begin{figure}
	\begin{center}
		\includegraphics[width=1.0\textwidth]{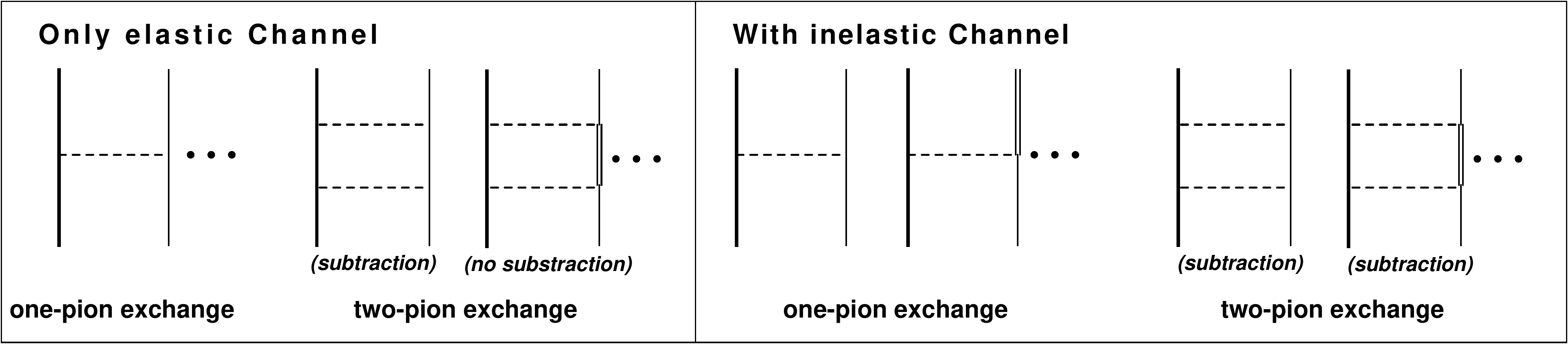}
	\end{center}
	\caption{Two approaches to couple-channel effect.}
	\label{fig:cc}
\end{figure}
\begin{figure}[htbp]
	\begin{minipage}[b]{0.37\textwidth}
		\centering
		\includegraphics[width=1\textwidth]{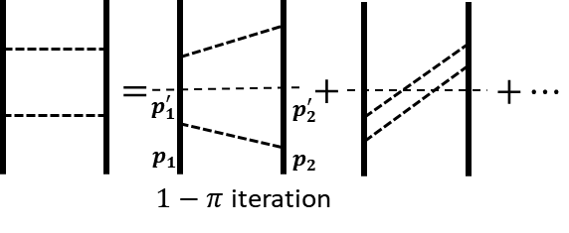}
		\caption{Box diagrams in TOPT.}\label{fig:box}
	\end{minipage}~~~~
	\begin{minipage}[b]{0.6\textwidth}
			\includegraphics[width=1\textwidth]{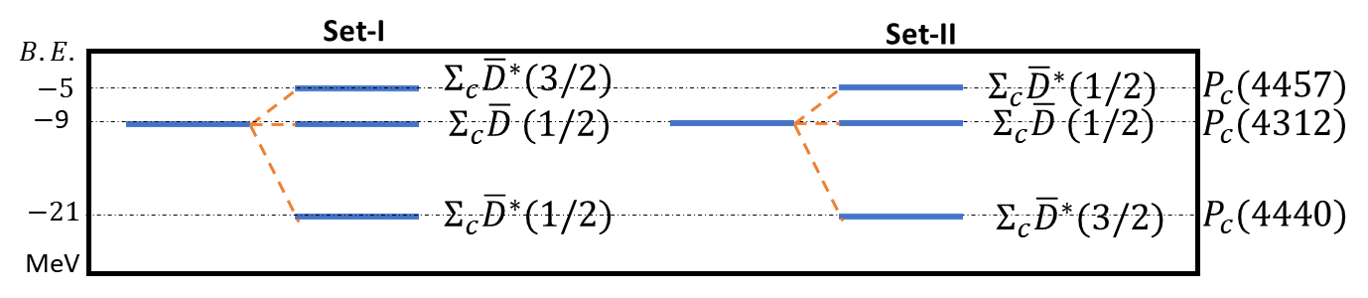}
		\caption{Two possible spin assignments for $P_c$ states.}\label{fig:spin}
	\end{minipage}
\end{figure}

For the box diagrams, we could deal with the couple-channel effect through two approaches as illustrated in Fig.~\ref{fig:cc}. From the view of time-ordered perturbation theory (TOPT), the box diagrams can be divided into the one-pion-iteration part and other part, see Fig.~\ref{fig:box}. For the one-pion-iteration part, the amplitude reads,
\begin{eqnarray}
\mathcal{M}&\sim&{(E_{\mathrm{inital}}-E_{\mathrm{inter}})}^{-1} \nonumber\\
&\approx&\left[\left(m_{1}+m_{2}+\frac{\bm{p}_{1}^{2}}{2m_{1}}+\frac{\bm{p}_{2}^{2}}{2m_{2}}\right)-\left(m'_{1}+m'_{2}+\frac{\bm{p}_{1}^{\prime2}}{2m^{\prime}_{1}}+\frac{\bm{p}_{2}^{\prime2}}{2m^{\prime}_{2}}\right)\right]^{-1}.
\end{eqnarray}
\iffalse
where $E_{\mathrm{inital}}$ and $E_{\mathrm{inter}}$ are the energies of the initial and intermediate states, respectively. $m_i(m_i^\prime)$ and $\bm{p}_i(\bm{p}_i^\prime)$ are the mass and 3-momentum of the initial (intermediate) states, respectively.
\fi
When the intermediate states are the same as the external lines, i.e., $m_1=m_2$ and $m'_1=m'_2$, like $NN$ systems, the amplitude would be amplified to destroy the power counting. We adopt the Weinberg's formalism to handle this problem~\cite{Weinberg:1991um}. We subtract the one-pion-iteration part in the box diagrams and recover their contribution by solving the Lippmann-Schwinger or Schr\"odinger equation. For $\Sigma_{c}^{(*)}\bar{D}^{(*)}$ systems, the intermediate states could be different from the external lines. The power counting works well for these box diagrams. Therefore, we could either subtract the one-pion-iteration part and include inelastic channels in tree diagrams or keep the one-pion-iteration part and only include the elastic channels in tree diagrams. In this work, we adopt latter approaches as shown in the left panel of Fig.~\ref{fig:cc}.

\section{Results and discussions}
Our LO results keep the heavy quark symmetry, which read,
\begin{equation}
{\cal
	V}_{\Sigma_{c}\bar{D}}=V_c,\quad
{\cal V}_{\Sigma_{c}\bar{D}^*}=V_c+V_{ss}\bm{S}_1\cdot\bm{S}_2.
\end{equation}
There are two possible spin assignments for the three $P_c$ states as shown in Fig.~\ref{fig:spin}. Three states have the same central potential and their binding energy splittings arise from the spin-spin interaction. If we treat spin-spin interaction as a perturbation, the ratio of the binding energy splitting is exactly the ratio of spin-spin interaction. Therefore, the spin Set-I in Fig.~\ref{fig:spin} is favored.

At the NLO, we include loop diagrams which bring some novel effects. The mass splittings between HQS partner states give rise to considerable HQSS violation for $\Sigma_c^{(*)}\bar{D}$ systems. Meanwhile, the new spin structure $(\bm{S}_1\cdot\bm{S}_2)^2$ appears at the loop level for the $\Sigma_{c}^*\bar{D}^*$~\cite{Wang:2019ato}, which is different from the two nucleon systems.

\begin{figure}
	\begin{center}
		\includegraphics[width=0.45\textwidth]{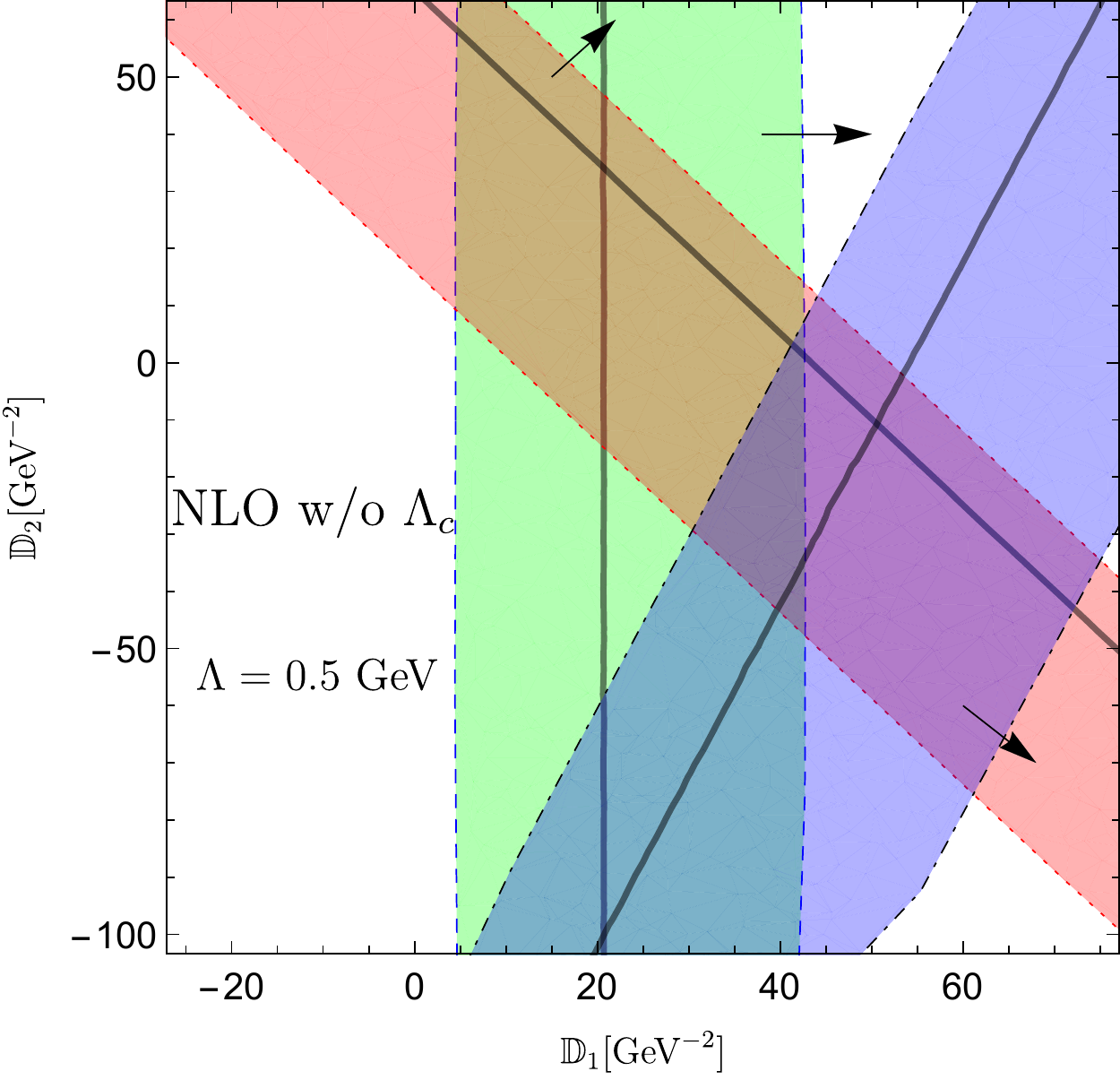}
		\includegraphics[width=0.45\textwidth]{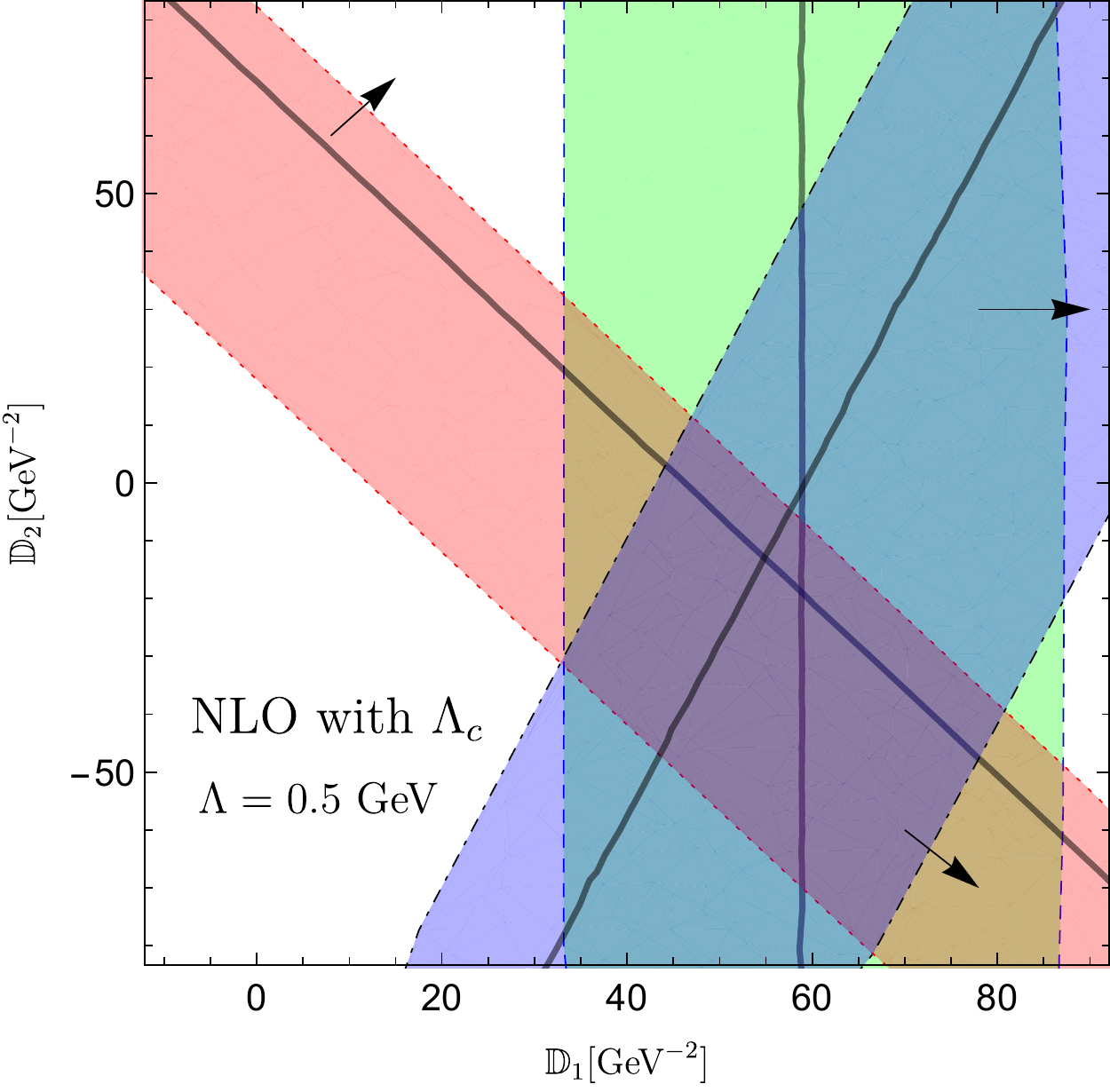}
	\end{center}
	\caption{The numerical results without and with $\Lambda_{c}$ intermediate states to the NLO.}
	\label{fig:results}
\end{figure}

We have two unknown low energy constants (LECs), $\mathbb{D}_1$ and $\mathbb{D}_2$, to the NLO. We vary the two LECs and try to reproduce three states simultaneously. The result is presented in Fig.~\ref{fig:results}. Three bands correspond to the regions of parameters in which the bound states have the binding energy $[-30,0]$ MeV. The three black lines represent the set of parameters corresponding to the experimental results. The left and right graphs in Fig.~\ref{fig:results} are the results without and with the $\Lambda_{c}$ as intermediate states, respectively. The three $P_c$ states can coexist only when the couple-channel effect from $\Lambda_{c}\bar{D}^{(*)}$ is considered~\cite{Wang:2019ato}. With the $\mathbb{D}_1$ and $\mathbb{D}_2$ being fixed, there exist bound solutions for all seven $S$-wave $\Sigma_{c}^{(*)}\bar{D}$ channels as listed in Table.~\ref{tab:binding}. It is possible to find the predicted states in the future experiments.
\begin{table}
	\tbl{The binding energies $\Delta E$ of $I=\frac{1}{2}$ hidden-charm $\Sigma_{c}^{(*)}\bar{D}^{(*)}$ systems.}
	{\begin{tabular}{cccccccc}
			\toprule
			&$[\Sigma_c\bar{D}]_{\frac{1}{2}}$&$[\Sigma_c\bar{D}^\ast]_{\frac{1}{2}}$&$[\Sigma_c\bar{D}^\ast]_{\frac{3}{2}}$&$[\Sigma_c^\ast\bar{D}]_{\frac{3}{2}}$&$[\Sigma_c^\ast\bar{D}^\ast]_{\frac{1}{2}}$&$[\Sigma_c^\ast\bar{D}^\ast]_{\frac{3}{2}}$&$[\Sigma_c^\ast\bar{D}^\ast]_{\frac{5}{2}}$\\
			\colrule
			$\Delta E$ &$-4.6$&$-22.5$&$-3.2$&$-34.5$&$-14.3$&$-3.4$&$-0.3$\\
			\botrule
		\end{tabular}
	}\label{tab:binding}
\end{table}
\begin{table}
	\tbl{Comparisons of the $\Sigma_{c}^{(*)}\bar{D}^{(*)}$ and $N$-$N$ systems.}
	{\begin{tabular}{lcc}
			\toprule
			& $N$-$N$ & $\Sigma_{c}^{(*)}\bar{D}^{(*)}$\\
			\colrule
			Chiral dynamics & $N-N$ & $\text{ (light diquark)-(light quark})$\\
			Inter. states & $\Delta$ & $\Lambda_{c}$\\
			\multirow{2}{*}{Inter. states} & \multirow{2}{*}{-} & HQS partner states \\
			&  & e.g., $c(qq)_{s=1}^{I=1}=\Sigma_{c}+\Sigma_{c}^{*}$\\
			\multirow{2}{*}{Heavy quark limit} & \multirow{2}{*}{-} & Pinch singularity in elastic channel\\
			&  & Including inelastic channel\\
			$M_{\mathrm{inter}}$ vs $M_{\mathrm{initial}}$ & $M_{\Delta}>M_{N}$ & e.g., $M_{\Lambda_{c}}+M_{\pi}<M_{\Sigma_{c}}$\\
			Imaginary part & w/o & w/\\
			Small scale expansion & works & fails in $\Sigma_{c}\bar{D}\ensuremath{-}\Lambda_{c}\bar{D}^{*}\ensuremath{-}\Sigma_{c}\bar{D}$\\
			Spin structure & $\bm{\sigma_{1}\cdot\sigma_{2}}$ & $\bm{S_{1}\cdot S_{2}},\quad(\bm{S_{1}\cdot S_{2}})^{2}...$\\
			Weinberg composite & $|E_{D}|\ll {m_{\pi}^2\over 2\mu}$& $|E_{P_c}|\sim {m_{\pi}^2\over 2\mu}\approx 9 \text{ MeV}$  \\\botrule
		\end{tabular}
	}\label{tab:compare}
\end{table}

In Table~\ref{tab:compare}, we compare the $\Sigma_{c}^{(*)}\bar{D}^{(*)}$ systems with two nucleon systems.  $\Sigma_{c}^{(*)}\bar{D}^{(*)}$ interactions have some interesting features which do not appear in $N$-$N$ systems. For example, the chiral dynamics for $\Sigma_{c}^{(*)}\bar{D}^{(*)}$ systems is the interplay between the light-light diquark in the charmed baryons and the light quark in the charmed mesons. Thus, one should take all the heavy quark partner states of charmed hadrons into consideration to include the whole interacting elements.

Our analytical results are pion mass dependent. The lattice QCD simulations for the $\Sigma_{c}^{(*)}\bar{D}^{(*)}$ systems are called for. Our expressions could be used to extrapolate the lattice QCD results to the physical pion mass~\cite{Liu:2016wxq}.

\section*{ACKNOWLEDGMENTS}
This project is supported by the National Natural Science Foundation of China under Grants 11575008,
11621131001.

% Unnumbered appendix sections can be obtained using \verb|\section*|.

\bibliographystyle{ws-procs9x6} % for numbered citation & references

\begin{thebibliography}{10}
%\cite{Aaij:2015tga}
\bibitem{Aaij:2015tga}
LHCb Collab. (R.~Aaij {\it et al.}),
%``Observation of $J/\psi p$ Resonances Consistent with Pentaquark States in $\Lambda_b^0 \to J/\psi K^- p$ Decays,''
{\it Phys. Rev. Lett.} {\bf 115}, 072001 (2015).
%doi:10.1103/PhysRevLett.115.072001
%[arXiv:1507.03414 [hep-ex]].
%%CITATION = doi:10.1103/PhysRevLett.115.072001;%%
%870 citations counted in INSPIRE as of 18 Nov 2019

%\cite{Chen:2016qju}
\bibitem{Chen:2016qju}
H.~X.~Chen, W.~Chen, X.~Liu and S.~L.~Zhu,
%``The hidden-charm pentaquark and tetraquark states,''
{\it Phys. Rept.}  {\bf 639}, 1 (2016)
%doi:10.1016/j.physrep.2016.05.004
%[arXiv:1601.02092 [hep-ph]].
%%CITATION = doi:10.1016/j.physrep.2016.05.004;%%
%461 citations counted in INSPIRE as of 18 Nov 2019



%\cite{Aaij:2019vzc}
\bibitem{Aaij:2019vzc}
LHCb Collab. (R.~Aaij {\it et al.}),
%``Observation of a narrow pentaquark state, $P_c(4312)^+$, and of two-peak structure of the $P_c(4450)^+$,''
{\it Phys. Rev. Lett.}  {\bf 122}, 222001 (2019).
%doi:10.1103/PhysRevLett.122.222001
%[arXiv:1904.03947 [hep-ex]].
%%CITATION = doi:10.1103/PhysRevLett.122.222001;%%
%99 citations counted in INSPIRE as of 18 Nov 2019



%\cite{Chen:2019asm}
\bibitem{Chen:2019asm}
R.Chen, Z.F.Sun, X.Liu and S.L.Zhu,
%``Strong LHCb evidence supporting the existence of the hidden-charm molecular pentaquarks,''
{\it Phys. Rev. D} {\bf 100}, 011502 (2019).
%doi:10.1103/PhysRevD.100.011502
%[arXiv:1903.11013 [hep-ph]].
%%CITATION = doi:10.1103/PhysRevD.100.011502;%%
%48 citations counted in INSPIRE as of 18 Nov 2019



%\cite{Liu:2019tjn}
\bibitem{Liu:2019tjn}
M.~Z.~Liu, Y.~W.~Pan, F.~Z.~Peng, M.~S\'anchez S\'anchez, L.~S.~Geng, A.~Hosaka and M.~Pavon Valderrama,
%``Emergence of a complete heavy-quark spin symmetry multiplet: seven molecular pentaquarks in light of the latest LHCb analysis,''
{\it Phys. Rev. Lett.}  {\bf 122}, 242001 (2019).
%doi:10.1103/PhysRevLett.122.242001
%[arXiv:1903.11560 [hep-ph]].
%%CITATION = doi:10.1103/PhysRevLett.122.242001;%%
%52 citations counted in INSPIRE as of 18 Nov 2019



%\cite{Guo:2019fdo}
\bibitem{Guo:2019fdo}
F.~K.~Guo, H.~J.~Jing, U.~G.~Mei{\ss}ner and S.~Sakai,
%``Isospin breaking decays as a diagnosis of the hadronic molecular structure of the $P_c(4457)$,''
{\it Phys. Rev. D} {\bf 99}, 091501 (2019).
%doi:10.1103/PhysRevD.99.091501
%[arXiv:1903.11503 [hep-ph]].
%%CITATION = doi:10.1103/PhysRevD.99.091501;%%
%40 citations counted in INSPIRE as of 18 Nov 2019



%\cite{Du:2019pij}
\bibitem{Du:2019pij}
M.~L.~Du, V.~Baru, F.~K.~Guo, C.~Hanhart, U.~G.~Mei{\ss}ner, J.~A.~Oller and Q.~Wang,
%``Evidence that the LHCb ${P_c}$ states are hadronic molecules and the existence of a narrow $P_c(4380)$,''
arXiv:1910.11846.
%%CITATION = ARXIV:1910.11846;%%



%\cite{Epelbaum:2008ga}
\bibitem{Epelbaum:2008ga}
E.~Epelbaum, H.~W.~Hammer and U.~G.~Mei{\ss}ner,
%``Modern Theory of Nuclear Forces,''
{\it Rev. Mod. Phys.}  {\bf 81}, 1773 (2009).
%doi:10.1103/RevModPhys.81.1773
%[arXiv:0811.1338 [nucl-th]].
%%CITATION = doi:10.1103/RevModPhys.81.1773;%%
%1029 citations counted in INSPIRE as of 18 Nov 2019



%\cite{Machleidt:2011zz}
\bibitem{Machleidt:2011zz}
R.~Machleidt and D.~R.~Entem,
%``Chiral effective field theory and nuclear forces,''
{\it Phys. Rept.}  {\bf 503}, 1 (2011).
%doi:10.1016/j.physrep.2011.02.001
%[arXiv:1105.2919 [nucl-th]].
%%CITATION = doi:10.1016/j.physrep.2011.02.001;%%
%775 citations counted in INSPIRE as of 18 Nov 2019



%\cite{Meng:2019ilv}
\bibitem{Meng:2019ilv}
L.~Meng, B.~Wang, G.~J.~Wang and S.~L.~Zhu,
%``The hidden charm pentaquark states and $\Sigma_c\bar{D}^{(*)}$ interaction in chiral perturbation theory,''
{\it Phys. Rev. D} {\bf 100}, 014031 (2019).

%%CITATION = doi:10.1103/PhysRevD.100.014031;%%
%15 citations counted in INSPIRE as of 18 Nov 2019



%\cite{Wang:2019ato}
\bibitem{Wang:2019ato}
B.~Wang, L.~Meng and S.~L.~Zhu,
%``Hidden-charm and hidden-bottom molecular pentaquarks in chiral perturbation theory,''
arXiv:1909.13054.
%%CITATION = ARXIV:1909.13054;%%
%2 citations counted in INSPIRE as of 18 Nov 2019



%\cite{Weinberg:1990rz}
%\bibitem{Weinberg:1990rz}
%S.~Weinberg,
%``Nuclear forces from chiral Lagrangians,''
%{\it Phys. Lett. B} {\bf 251}, 288 (1990).
%doi:10.1016/0370-2693(90)90938-3
%%CITATION = doi:10.1016/0370-2693(90)90938-3;%%
%1226 citations counted in INSPIRE as of 18 Nov 2019



%\cite{Weinberg:1991um}
\bibitem{Weinberg:1991um}
S.~Weinberg,
%``Effective chiral Lagrangians for nucleon - pion interactions and nuclear forces,''
{\it Nucl. Phys. B} {\bf 363}, 3 (1991).
%doi:10.1016/0550-3213(91)90231-L
%%CITATION = doi:10.1016/0550-3213(91)90231-L;%%
%1170 citations counted in INSPIRE as of 18 Nov 2019

%\cite{Liu:2016wxq}
\bibitem{Liu:2016wxq} 
Z.~W.~Liu, J.~M.~M.~Hall, D.~B.~Leinweber, A.~W.~Thomas and J.~J.~Wu,
%``Structure of the $\mathbf{\Lambda(1405)}$ from Hamiltonian effective field theory,''
{\it Phys.\ Rev.\ D {\bf 95}}, 014506 (2017)
%doi:10.1103/PhysRevD.95.014506
%[arXiv:1607.05856 [nucl-th]].
%%CITATION = doi:10.1103/PhysRevD.95.014506;%%
%23 citations counted in INSPIRE as of 21 Nov 2019



	
	
\end{thebibliography}

\end{document}